\documentclass[12pt,a4paper,onecolumn]{article}

\usepackage[T1]{fontenc}
\usepackage[utf8]{inputenc}
\usepackage{graphicx}
\usepackage{authblk}

\title{Formation of normal surface plasmon modes in small sodium nanoparticles.}
\author[1,2]{N.L. Matsko}

\affil[1]{\footnotesize Moscow Institute of Physics and Technology, Dolgoprudny, Moscow Region, 141701, Russia}
\affil[2]{P.N. Lebedev Physical Institute, Russian Academy of Sciences, Leninskii prosp. 53, 119991 Moscow, Russia}

\date{}

\begin{document}

\maketitle

\begin{abstract}
Formation of surface plasmon modes in sodium nanoclusters containing 20-300 atoms was studied using the GW method. It is shown that in the small Na nanoparticles up to 2 nm in size, the loss function $Im[\epsilon^{-1}]$ is dominated by a single peak corresponding to localized surface plasmon resonance (LSPR). For particles of 2 nm and more, a resonance corresponding to surface plasmon polariton (SPP) oscillations begins to form, as well as a resonance corresponding to volume plasmon (VP) excitations. Considering the above, the linear size of a particle in the range of 0.7-3.7 nm can be estimated as the lower limit for metal nanodevices operating with SPP. On the example of spherical nanoparticles consisting of a silicon core coated with sodium atoms, it is shown that the LSPR mode is selectively suppressed while the SPP mode is not. Such composite structures can be considered as an example of nanoplasmonic devices with selectively tuned characteristics.
\end{abstract}

\section{Introduction}
Prospects for the use of nanoscale plasmonic devices in industry promise a breakthrough in technology. Plasma excitations in nanoparticles have been actively investigated for several decades. By present time a large amount of experimental data related to plasmons in metal and semiconductor nanoparticles of various shapes and sizes has been accumulated. Theoretical works on calculating the characteristics of plasma excitations at the nanoscale are regularly published. Growing possibilities of numerical calculations and the possibility of taking into account various effects at a more accurate level contribute to it. The picture of collective electronic oscillations in nanoobjects is often non-trivial. It is explained by the presence of various plasmon modes, their interaction with each other, with the surface of the object and with the quasiparticle spectrum.

In general, the picture of plasma oscillations in finite systems can be represented as follows. Peaks in the nanoparticle absorption and scattering spectra \cite{mulvaney,yasumatsu,jaensch,link,kafle,tiggesbaumker,pollack,rod_fer,kumbhar}, in the EELS spectra for nanoparticles \cite{mitome,nienhaus,nakashima,wang} indicate several types of excitations. The high-frequency peak corresponds to volume plasmons with a frequency tending to $\omega_{pl}$ in the limit of large particles, where $\omega_{pl}$ is the frequency of the volume plasmon in the bulk. At a lower frequency, peaks corresponding to multipole plasma excitations are present. These peaks appear due to the closed surface of the particle. In the Mie theory \cite{mie}, these peaks correspond to frequencies equal to $\sqrt{\frac{l}{2l+1}}\omega_{pl}$. In particular, excitations with $l=1$ correspond to LSPR with the Mie resonance frequency equal to $\omega_{pl}/ \sqrt 3$. In addition, peaks corresponding to the normal surface eigenmodes are observed at the boundary of bulk samples. One of them is the SPP mode \cite{zayats1, zayats2}, which is an electron density wave propagating along the surface with a frequency of $\omega_{pl}/ \sqrt 2$ (for the jellium-vacuum interface). The induced density is localized in the surface layer and has a monopole structure normal to the interface. At a higher frequency (but less than $\omega_{pl}$), a resonance associated with the surface Bennett mode may be observed \cite{bennett,tsuei}. The Bennett mode has a dipole structure of induced charge density normal to the surface.

Most of the theoretical works related to plasma excitations in metal nanoobjects are devoted to the study of the dependence of the  LSPR mode frequency and attenuation on the particle size. To date, this issue has been studied in detail. There are many experimental and theoretical works considering the formation of quadrupole, octupole and other multipole modes in a nanocluster. At the same time, very few works consider normal surface plasmon modes in nanoclusters.
Such a gap is mainly due to the lack of experimental data on SPPs in nanoparticles. Direct excitation of SPP in a nanoparticle by an incident electromagnetic wave is not possible, since both the frequency and the wavevector of the exciting light should match the frequency and wavevector of the SPP \cite{maier,novotny,compaijen}. The more sophisticated method is needed. As the consequence related problems were not addressed in the theory. We do not know the microscopic theoretical works devoted to the mechanism of the formation and conditions of existence of such modes in small nanoparticles. The purpose of our work is to fill this gap. It was studied the behavior of SPP excitations in small nanoparticles. The limit of a nanoobject size where these modes exist as distinctive excitations was examined. Such research seems to be a very relevant topic not only from a fundamental point of view, but also from practical applications in perspective electronic devices \cite{zayats2,barnes,gramotnev}.

The choice of sodium nanoparticles as an object of study is explained by the following. On the one hand, a lot of data on the LSPR in sodium nanoparticles is available. On the other hand, sodium systems are relatively simple for the band description. Thus, it is possible to carry out accurate calculations of objects containing up to several hundred atoms, to compare the results with the available data, as well as to complement the existing picture of plasmons in nanoobjects.

\section{Computational details}

For accurate description of electronic excitations it is necessary to calculate the response of the system to a perturbation, which requires consideration in the framework of "beyond-DFT" methods. The most common method for studying plasmons in nanoparticles is the real-time TDDFT method \cite{durante,li,towsend,iida,lerme,xiang,stella}. Frequency-space TDDFT \cite{ekardt} and GW \cite{bulgac} approximations are less commonly used. In the real-time TDDFT, the system in the ground state is first affected by an appropriately selected external perturbation, and then the evolution of the system is calculated. The peculiarity of this method that needs to be emphasized is that the applied perturbation interacts with different plasma modes in different ways. For example, a perturbation resonantly exciting localized surface plasmons will not excite bulk ones. The activation of bulk modes will occur through their interaction with surface modes. Thus, the excitation of different modes is differently distributed over time. This seems to be the reason that the peaks of bulk plasmons in real-time TDDFT are often almost invisible. In the case of excitation of the SPP mode in nanoparticles, real-time TDDFT approximation will have the same problems. There is no such problem for the method based on GW approximation that is used in this work.

The approach is as follows. The dynamic response function of the system at the frequency $\omega$ has the form \cite{deslippe},\cite{hyblouie}:

\begin{equation} \label{f_1} \chi_{GG'}(q,\omega)=\sum_n^{occ}\sum_{n'}^{emp}\sum_k <n,k+q|e^{i(q+G)r}|n',k><n',k|e^{-i(q+G')r}|n,k+q>\end{equation}

\begin{equation}  \label{f_1} \times \frac{1}{2} \left[\frac{1}{E_{nk+q}-E_{n'k}-\omega-i\delta}+\frac{1}{E_{nk+q}-E_{n'k}+\omega+i\delta} \right]
\end{equation}

\centerline{}

where $q$ is a vector in the first Brillouin zone, $G$ and $G'$ are reciprocal-lattice vectors, $E_{nk}$ is the mean-field quasiparticle energy, $n$ and  $n'$ are occupied and unoccupied electronic states respectively. Based on the response function $\chi_{GG'}(q,\omega)$, a dielectric matrix is constructed:

\begin{equation} \label{f_2} \epsilon_{GG'}(q,\omega)=\delta_{GG'}-v(q+G)\chi_{GG'}(q,\omega)
\end{equation}

Then the inverse matrix $\epsilon^{-1}_{GG'}(q,\omega)$ is build. Finally, we construct $\epsilon^{-1}(r,r',\omega)$:

\begin{equation} \label{f_3} \epsilon^{-1}(r,r',\omega)=\sum_{q,G,G'} e^{i(q+G)r}\epsilon^{-1}_{GG'}(q,\omega)e^{-i(q+G')r'} \end{equation}

Calculations were carried out at the gamma point $q=0$. The function $\epsilon^{-1}(r,r',\omega)$ describes the system response at $r$ to a perturbation at $r'$. In the absence of local field effects in a homogeneous crystal (as well as in the jellium model) $\epsilon^{-1}(r,r')=\epsilon^{-1}(|r-r'|)$.
The required modes of collective electronic excitations are manifested as the corresponding peaks of the loss function $Im[\epsilon^{-1}]$ \cite{pines},\cite{sturm},\cite{haque},\cite{hanke}. For small clusters, we can distinguish $\epsilon^{-1}(r,r')$ for two cases, when $r$ and $r'$ are inside the cluster or at the surface region. In the first case, VP's should give a more pronounced contribution, in the second - SPP's.

Density functional calculations in the Quantum Espresso (QE) \cite{qe} code were used as a starting point for the one-iteration G$_0$W$_0$. The QE calculations were made with PBE GGA pseudopotential and a plane wave basis set having the cutoff energy of 45 Ry. Computations were performed for cubic supercell geometry with the side length of 62 Bohr. Such supercell sizes provide the vacuum region enough for the plasma frequencies convergence to within 0.1 eV.
For the GWA calculations the BerkeleyGW \cite{deslippe},\cite{hyb-Lou},\cite{bgw2} package was applied. Full frequency dependence method with contour-deformation formalism for the inverse dielectric matrix calculations was used. Energy cutoff for the dielectric matrix was set to 2.0 Ry. The Coulomb interaction was cutoff on the edges of cell box.

Sodium cluster geometries studied in our work were taken from The Cambridge Cluster Database \cite{cambridge}. The atomic structures of considered nanoclusters were relaxed using PBE GGA functional until atomic forces became less than $10^{-4}$ Ry/\AA. The resulting structures have a compact shape close to spherical. The internal structure of clusters containing more than one hundred Na atoms tends to the bcc geometry of crystalline sodium.
However the exact atomic structure of the nanoparticle is not crucial. Since plasmons in metal clusters are the collective motion of conduction electrons, their general properties are rather related to the overall shape, size, and electronic density than to the exact atomic structures \cite{durante},\cite{li},\cite{reinhard}. This suggests that the studied structures give an idea of the characteristic properties of plasma excitations in metallic nanoparticles.

Si$_{n}$Na$_{m}$ structures were obtained by coating a compact silicon core \cite{myplasma} with a shell of sodium atoms with subsequent relaxation. For clarity, calculations of the plasma excitations of the structure obtained from Na$_{300}$ by cutting the core out (only the outer layer of atoms is left) were performed. Obviously, such a structure is not stable, there was no relaxation in it.

\section{Results and discussion}

\begin{figure}[h]
\centering
\includegraphics[width=0.4\textwidth]{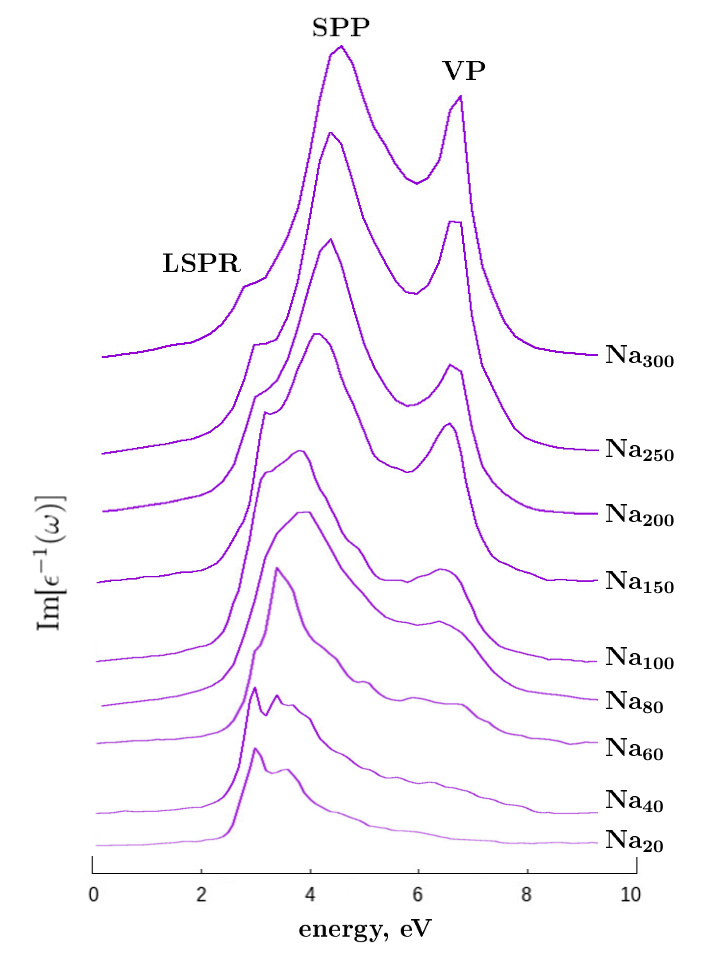}
\caption{The loss function $Im[\epsilon^{-1}(\omega)]$ for sodium clusters containing 20-300 atoms. Peaks corresponding to LSPR, SPP and VP excitations are marked for the upper curve, i.e. for the cluster Na$_{300}$.}
\label{fig1}
\end{figure}

Let us consider how the resonances of collective electronic excitations change with an increase of the nanoparticle, starting with structures containing several tens of atoms. For small clusters, it is reasonable to expect smearing of plasmon resonances due to strong attenuation during plasmon scattering on the surface \cite{mitome,nienhaus,wang} and the influence of single-particle excitations \cite{li,sturm,bohm,nesterenko,xia}. The significance of these effects should decrease with increasing cluster size. Figure 1 shows the loss function $Im[\epsilon^{-1}(r,r',\omega)]$ for sodium clusters containing 20, 40, 60, 80, 100, 150, 200, 250, 300 atoms. The point $r$ was chosen at the center of the particle, and the point $r'$ was chosen near the surface. In the Na$_{20}$-Na$_{60}$ clusters a single resonance in the region of 3-3.5 eV can be distinguished, it corresponds to the LSPR. This resonance is highly blurred and has a pronounced tail extending to the high-frequency region. Such structure of LSPR in small Na nanoparticles is explained by the Landau damping due to the presence of single-particle transitions with energies close to the LSPR energy \cite{li,sturm,bohm,nesterenko,xia}. As can be seen from fig.1, among the nanoparticles considered, the highest peak defragmentation is observed in Na$_{20}$ and Na$_{40}$, which corresponds to the experiment and was noted in previous theoretical works \cite{pollack,li,bulgac,xia}.

As the cluster size increases (Na$_{80}$-Na$_{100}$), two more peaks associated with SPP and VP resonances begin to form at frequencies 4-4.6 eV and 6.5-6.8 eV, respectively. This identification of peaks with resonances can be justified as follows. The resonance intensity in the region of 6.5-6.8 eV (resonance associated with VPs) increases with the cluster size (fig.1) and is maximum inside a particle (fig.2), as it should be in the case of bulk plasmons. The VP frequency in crystalline sodium is 5.8-5.9 eV \cite{haque},\cite{vomfelde},\cite{schnatterly}. Our calculation for VP frequency in sodium gives a magnitude of 6.05 eV that is typical overestimation for the RPA models \cite{tsuei}. The VP frequency in nanoparticles is increased compared to the bulk, which is explained by quantum size effect \cite{xiang,ekardt,sonnichsen,ouyang}. Thus, the resonance in the region of 6.5–6.8 eV can be identified as VP oscillations. From fig.1 it is seen that in clusters containing 100-150 atoms, the intensity of VPs becomes comparable to surface plasmons intensity. In the work \cite{ekardt}, VP resonance also becomes distinguishable for a jellium sphere containing about 100 electrons. In the work \cite{bystryi}, VP mode becomes clearly distinguishable for ionized sodium nanoclusters containing about 10000 atoms or more, which is probably the result of non-stationary and non-linear effects in high temperature plasma.

The resonance intensity in the region of 4–4.6 eV (resonance associated with SPPs) is maximum in the surface region, which corresponds to surface excitations (fig.2). The frequency of the peak increases with increasing particle size and for Na$_{300}$ reaches 4.6 eV, i.e. it approaches the value $\omega_{pl}/ \sqrt 2$ ($\omega_{pl}$= 6.8 eV for Na$_{300}$), as it should be for the surface of the bulk system. This allows associating the peak with the SPP resonance.

\begin{figure}[h]
\centering
\includegraphics[width=0.4\textwidth]{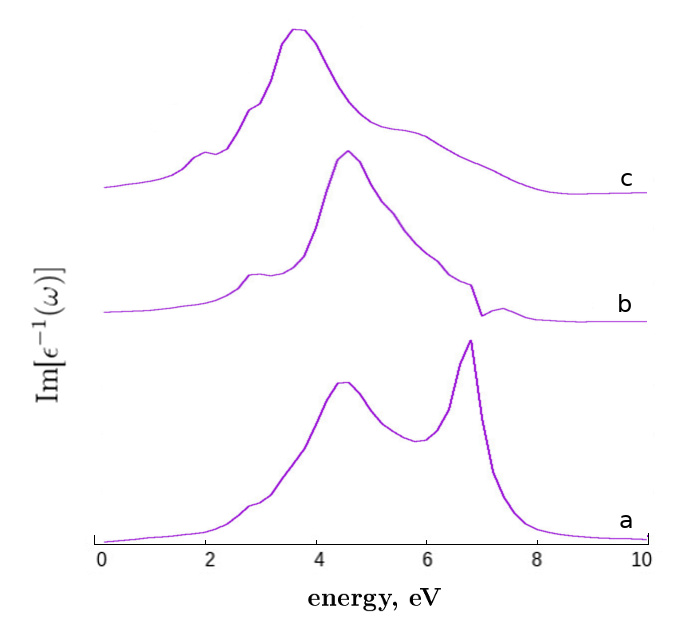}
\caption{The loss function $Im[\epsilon^{-1}(r,r',\omega)]$ in Na$_{300}$. {\bf a} - points $r$ and $r'$ are in the central region of the particle, {\bf b} - points $r$ and $r'$ are in the surface region. {\bf c} - the loss function for the structure obtained from Na$_{300}$ by cutting the core out (only the outer layer of atoms is left).}
\label{fig2}
\end{figure}

In Na$_{80}$ and Na$_{100}$ LSPR and SPP peaks are still difficult to separate (fig.1). With the further increase of the particle, these resonances become detached. LSPR peak frequency decreases from the cluster Na$_{150}$ to Na$_{300}$ from 3.2 eV to 2.8 eV (see also \cite{xiang}). SPP resonance frequency increases from 4 eV for Na$_ {100}$ to 4.6 eV for Na$_{300}$. Summarizing, we can say that SPPs in Na nanoparticles become well distinguishable excitation for systems containing 150-200 atoms and more, or with a size of $\sim$ 2 nm and more.

Figure 2 shows the function $Im[\epsilon^{-1}(r,r',\omega)]$ for Na$_{300}$, where the points $r$ and $r'$ are in the central region of the particle (a), and where the points $r$ and $r'$ are in the surface region (b). It is clearly seen that in the first case, the highest intensity has a high-frequency resonance at 6.8 eV, corresponding to VPs. In the case of $r$ and $r'$ lying in the surface region, the peak at 4.6 eV corresponding to SPPs has the highest intensity. In both cases, the LSPR peak at a frequency of 2.8 eV is well distinguishable. Figure 2c shows the loss function for a nanocluster obtained by cutting out the inner part from Na$_{300}$, only atoms of the outer layer are left. The form of the loss function in this hollow cluster does not fundamentally change for different positions of $r$ and $r'$. As can be seen in this case, a very weak signal is visible at the site of the VP resonance, while the LSPR and SPP resonances are redshifted. This decrease in frequency is explained by the fact that the electron density of the shell monatomic layer is less than in the surface layer of a simple Na$_{300}$, since the electron density spills out both outward and inward. A similar redshift of the LSPR was observed experimentally in hollow gold nanoclusters \cite{liang}.

In some theoretical works (e.g. \cite{xiang},\cite{stella}), relatively weak features (relative to other excitations) in the optical absorption spectra have been attributed to the surface Bennett mode. In the electron loss spectra experiment, the Bennett mode resonance is observed on the bulk Na \cite{tsuei} surface in the middle between SPP and VP peaks. From fig.2c for the hollow particle we can see the tail in the spectrum of the function $Im[\epsilon^{-1}]$ at frequencies of 5-8 eV. A detailed study of this tail reveals two weak peaks at frequencies of 6 and 7 eV. If we attribute these peaks to the Bennett and the VP resonance respectively, then such a picture will generally correspond to the position of the resonances observed in the experiment for the bulk sodium surface.
If this matching is correct and the marked features are a manifestation of Bennett mode, it turns out that in a hollow particle such features are more noticeable. This could be related to the charge distribution of plasma oscillations in the monoatomic layer of the shell. For the charge oscillations perpendicular to surface a negative electron charge is collected on the one side of the atomic layer and a positive ion charge remains on the other. Thus, the condition of the dipole charge distribution for the mode is satisfied more efficiently.

Quadrupole plasmon excitations as well as multipole modes of higher orders were not considered in the analysis of the studied clusters, since such modes are excited in much larger particles. For example, in gold and silver nanoparticles quadrupole plasmon resonance is observed at particle sizes of $\sim$100 nm and more \cite{rod_fer},\cite{kumbhar},\cite{millstone},\cite{zhou},\cite{kolwas}.

\begin{figure}[h]
\centering
\includegraphics[width=0.45\textwidth]{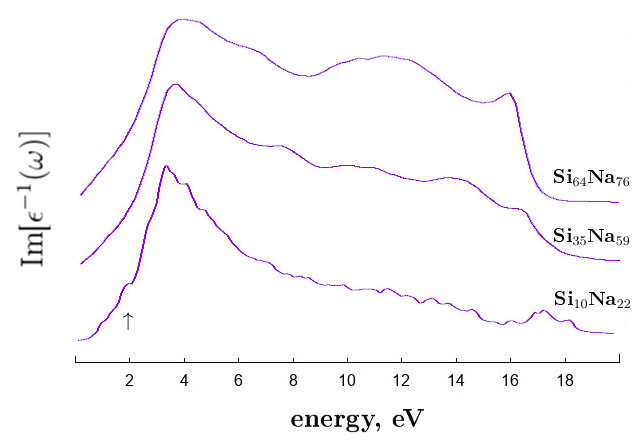}
\caption{The loss function in clusters Si$_{10}$Na$_{22}$, Si$_{35}$Na$_{59}$ and Si$_{64}$Na$_{76}$. Arrow marks LSPR peak in Si$_{10}$Na$_{22}$.}
\label{fig3}
\end{figure}

Figure 3 shows the function $Im[\epsilon^{-1}(r,r',\omega)]$ ($r$ and $r'$ lie on opposite half radii) for Si$_{10}$Na$_{22}$, Si$_{35}$Na$_{59}$ and Si$_{64}$Na$_{76}$ nanoclusters where the compact silicon core is coated with a shell of sodium atoms. The peculiarity of such a system is that silicon and sodium atoms do not form a chemical bond. From fig. in supplementary it can be seen that the electron density of the silicon core is concentrated around the silicon atoms and in the bonds between them. The electrons of the sodium shell are evenly distributed in the surface layer. In a rough approximation, we can assume that the electronic structure of such Si-Na nanoparticle is represented by two separate subsystems - the silicon core and the sodium shell. Then the plasmons in these two subsystems can be considered as distinct and "weakly" interacting excitations. The characteristic frequencies of electronic excitations in these systems also differ markedly. For comparison, the frequencies of bulk plasmons in crystalline sodium and crystalline silicon are 5.8-5.9 eV \cite{haque},\cite{vomfelde},\cite{schnatterly} and 16.9-17.3 eV \cite{si_Wpl},\cite{si_Wpl2},\cite{si_Wpl_Spl}, respectively.

For fig.3, the following features can be pointed. In the frequency region of 18–14 eV, the studied clusters exhibit peaks associated with the bulk plasmons of the silicon core \cite{mitome},\cite{nienhaus},\cite{myplasma}. Peaks in the region of 13–7 eV are associated with the surface modes of the silicon core. The highest intensity on all the curves has the SPP peak of the sodium shell at a frequency of 3.4, 3.7 and 4 eV for the clusters Si$_{10}$Na$_{22}$, Si$_{35}$Na$_{59}$ and Si$_{64}$Na$_{76}$ respectively. The absence of the LSPR peak of the sodium shell in Si$_{35}$Na$_{59}$ and Si$_{64}$Na$_{76}$ is of particular interest. This can be explained by the fact that LSPR is a uniform translational oscillations of electrons relative to the center of mass of the particle. The silicon core screens the Coulomb interaction (with a frequency lower than the plasma frequency of the core) of ordered electron displacement of the LSPR mode at opposite ends of the sodium shell. In the case of a small silicon core as in Si$_{10}$Na$_{22}$, the LSPR peak at a frequency of 2 eV is still distinguishable, but is strongly suppressed. The LSPR in sodium shell disappears as the size of the silicon core increases (Si$_{35}$Na$_{59}$ and Si$_{64}$Na$_{76}$). In turn, SPP excitation is a charge density wave propagating along the surface, so the central core affects the SPP resonance of the metal shell of the particle much less.
Unfortunately, the comparing of the obtained results for Si-Na clusters with experimental data seems problematic. There are a few papers on optical absorption in the metal coated semiconductor nanoparticles \cite{zhouhaus,zhouhaus2,lee}, but the nanoparticles studied in these papers are two orders of magnitude larger than the particles discussed here, and the effect of the LSPR mode screening cannot be distinguished against the background of other effects.

\section{Conclusions}

The above calculations show that in the small Na nanoparticles up to 2 nm in size and containing up to a hundred atoms, the loss function $Im[\epsilon^{-1}]$ demonstrates a single resonance corresponding to LSPR. When size of a particle reaches 2-2.5 nm (200-300 atoms) or more, a clearly distinguishable peak corresponding to SPPs begins to form in the function $Im[\epsilon^{-1}]$, the relative intensity of the LSPR decreases. A resonance corresponding to volume plasma excitation appears. Thus, from the point of view of collective electronic excitations, the inner and the surface regions are formed in the nanocluster. In the surface region, the electron density changes from zero value (at the boundary with the vacuum) to the value corresponding to the inner region. Note that the "surface region" at the bulk sodium boundary in the case of a static external field has a comparable thickness: the depth of localization of the induced charge (with the structure of Friedel oscillations) is about 1.5-2 nm \cite{liebsch}. Considering the above, we conclude that the linear size in the region of 2 nm is the lower limit for sodium-based devices operating with SPP. For nanoparticles of other metals or alloys, the minimum size should vary proportionally to the plasma screening length in the material. The Debye length could be expressed as $\lambda_D\sim \sqrt{r_s}^3$, where $r_s$ is Wigner radius. Given that in sodium $r_s$ equals to 4 a.u., and varies from 2 to 6 a.u. in general for metals \cite{ashkroft}, the lower limit for SPP devices will vary in the range of 0.7-3.7 nm.

Si$_{n}$Na$_{m}$ nanoparticles having a structure with a compact silicon core coated with a shell of sodium atoms were considered. Calculations show that the semiconductor core screens the Coulomb interaction of the LSPR mode electron displacements in the metal shell. As a result, in the considered structures Si$_{35}$Na$_{59}$ and Si$_{64}$Na$_{76}$ the LSPR mode in sodium shell is completely suppressed. At the same time, the semiconductor core has a much smaller effect on SPP excitation in the metal shell, SPP resonance remains clearly visible. Such combined structures can be considered as an example of nanoplasmonic devices with selective tuned features.

\section{Acknowledgements}

Calculations were carried out on the Joint Supercomputer Center of the Russian Academy of Sciences (JSCC RAS). The work was supported by the Russian Science Foundation (grant 19-72-30043).

\section{Supplementary}

\begin{figure}[h]
\centering
\includegraphics[width=0.9\textwidth]{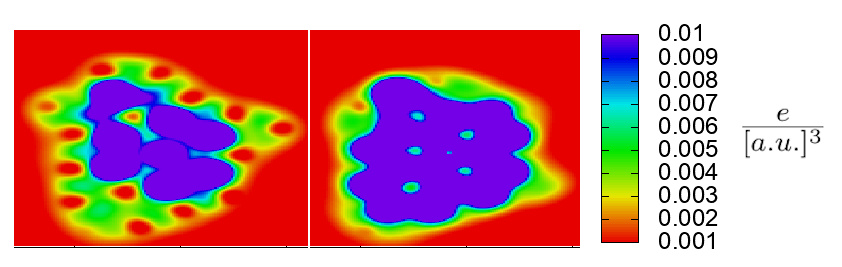}
\caption{Slices of the electron density of the Si$_{64}$Na$_{76}$ nanocluster, [e/a.u.]$^3$. The violet region is the silicon core, the yellow-green region is the electrons of the sodium shell.}
\label{fig4}
\end{figure}

\bibliography{thesis}

\begin{thebibliography}{99}
\bibitem{mulvaney} P. Mulvaney,  Langmuir, 12, 788-800 (1996)
\bibitem{yasumatsu} H. Yasumatsu, T. Kondow, H. Kitagawa, K. Tabayashi, and K. Shobatake, J. Chem. Phys. {\bf104}, 899 (1996).
\bibitem{jaensch} R. Jaensch and W. Kamke, Mol. Mater. {\bf13}, 143 (2000).
\bibitem{link} S. Link and M.A. El-Sayed, J. Phys. Chem. B {\bf103}, 8410-8426 (1999).
\bibitem{kafle} B.P. Kafle, H. Katayanagi, M. Prodhan, H. Yagi, C. Huang and K. Mitsuke, J. Phys. Soc. Japan. {\bf77}, 014302 (2008).
\bibitem{tiggesbaumker} J. Tiggesbaumker, L. Koller, K. H. Meiwes-Broer, and A. Liebsch, Phys.Rev. A {\bf48}, R1749 (1993).
\bibitem{pollack} S. Pollack, C.R.C. Wang, and M.M. Kappes, J. Chem. Phys. {\bf94}, 2496 (1991).
\bibitem{rod_fer} J. Rodriguez-Fernandez, J. Perez-Juste, F.J. Garcia de Abajo, and L.M. Liz-Marzan, Langmuir {\bf22}, 7007-7010 (2006)
\bibitem{kumbhar} A.S. Kumbhar, M.K. Kinnan, and G. Chumanov, J. AM. CHEM. SOC. {\bf127}, 12444-12445 (2005).
\bibitem{millstone} J.E. Millstone, S. Park, K.L. Shuford, L. Qin, G.C. Schatz, and C.A. Mirkin, J. AM. CHEM. SOC. 2005, 127, 5312-5313
\bibitem{mitome} M. Mitome, Y. Yamazaki, H. Takagi, and T. Nakagiri. Journal of Applied Physics {\bf72}, 812 (1992).
\bibitem{nienhaus} H. Nienhaus, V. Kravets, S. Koutouzov, C. Meier, and A. Lorke et al. J Vac. Sci. Technol. B {\bf24}, 1156 (2006).
\bibitem{nakashima} P.N.H. Nakashima, T. Tsuzuki, and A.W.S. Johnson, Journal of Applied Physics {\bf85}, 3, 1556 (1999).
\bibitem{wang} J. Wang, X. An, and Q. Li, R.F. Egerton, Applied Physics Letters {\bf86}, 201911 (2005).
\bibitem{mie} G. Mie, Ann. Phys. 330, 377—445 (1908)
\bibitem{zayats1} A.V. Zayats and I.I. Smolyaninov, Journal of Optics A: Pure and Applied Optics, Volume 5, Number 4 (2003).
\bibitem{zayats2} A.V.Zayats, I.I. Smolyaninov, A.A. Maradudin, Physics Reports, Volume 408, Issues 3–4 (2005).
\bibitem{bennett} A.J. Bennett, Phys. Rev. B {\bf1}, 203 (1970).
\bibitem{tsuei} K.-D. Tsuei, E.W. Plummer, A. Liebsch, E. Pehlke, K. Kempa, P. Bakshi, Surface Science {\bf247}, 2–3, 302-326 (1991)
\bibitem{maier} S.A. Maier and H.A. Atwater, JOURNAL OF APPLIED PHYSICS {\bf98}, 011101 (2005)
\bibitem{novotny} L. Novotny, B. Hecht. Principles of Nano-Optics, Cambridge University Press (2008).
\bibitem{compaijen} P.J. Compaijen, V.A. Malyshev and J. Knoester, Optics Express, {\bf23} 3, pp. 2280-2292 (2015)
\bibitem{barnes} W.L. Barnes, A. Dereux, T. W. Ebbesen, Nature, {\bf424}, 824–830 (2003)
\bibitem{gramotnev} D.K. Gramotnev and sergey I. Bozhevolnyi, Nature photonics, {\bf4}, 83-91, (2010)
\bibitem{durante} N. Durante, A. Fortunelli, M. Broyer, and M. Stener, J. Phys. Chem. C 2011, 115, 6277–6282
\bibitem{li} J.-H. Li, M. Hayashi, and G.-Y. Guo, Phys. Rev. B {\bf88}, 155437 (2013).
\bibitem{towsend} E. Townsend and G.W. Bryant, Nano Lett. 12,1,429-434 (2012)
\bibitem{iida} K. Iida, M. Noda, K. Ishimura, and K. Nobusada, J. Phys. Chem. A {\bf118}, 11317-11322 (2014)
\bibitem{lerme} J. Lerme, H. Baida, C. Bonnet, M. Broyer, E. Cottancin, A. Crut, P. Maioli, N. Del Fatti, F. Vallee, and M. PellarinJ. Phys. Chem. Lett. 2010, 1, 2922–2928
\bibitem{xiang} H. Xiang, X. Zhang, D. Neuhauser, G. Lu, J. Phys. Chem. Lett. {\bf5}, 1163-1169 (2014)
\bibitem{stella} L. Stella, P. Zhang, F.J. García-Vidal, A. Rubio, and P. Garcia-Gonzalez, J. Phys. Chem. C, 117, 17,8941-8949 (2013)
\bibitem{ekardt} W. Ekardt, Phys.Rev.B {\bf31}, 6360 (1985).
\bibitem{bulgac} A. Bulgac and C. Lewenkopf, EPL {\bf31}, 519 (1995)
\bibitem{deslippe} J. Deslippe, G. Samsonidze, D.A. Strubbe, M. Jain, M.L. Cohen, S.G. Louie, Comp. Phys. Comm. {\bf183}, 1269–1289 (2012)
\bibitem{hyblouie} M.S. Hybertsen and S.G. Louie, Phys. Rev. B {\bf35}, 5585 (1987)
\bibitem{pines} D. Pines, "Elementary excitations in solids", Benjamin, New York-Amsterdam (1963).
\bibitem{sturm} K. Sturm, Electron energy loss in simple metals and semiconductors, Advances in Physics, 31:1, 1-64 (1982)
\bibitem{haque} M.S. Haque \& K.L. Kliewer. Plasmon Properties in bcc Potassium and Sodium. Phys. Rev. B, 7{\bf6}, 2416–2430 (1973)
\bibitem{hanke} W. Hanke, Dielectric theory of elementary excitations in crystals, Advances in Physics, 27:2, 287-341 (1978)
\bibitem{qe} P. Giannozzi et al. J.Phys.:Condens.Matter {\bf21}, 395502 (2009).
\bibitem{hyb-Lou} M.S. Hybertsen and S.G. Louie, Phys. Rev. B {\bf34}, 5390 (1986).
\bibitem{bgw2} M. Rohlfing and S.G. Louie, Phys. Rev. B {\bf62}, 4927 (2000).
\bibitem{cambridge} http://www-wales.ch.cam.ac.uk/~wales/CCD/Si.html
\bibitem{reinhard} P.G. Reinhard and E. Suraud, Introduction to Cluster Dynamics (Wiley-VCH, Weinheim, 2004)
\bibitem{myplasma} N.L. Matsko, Phys. Chem. Chem. Phys., 20, 24933-24939 (2018)
\bibitem{bohm} Bohm, D.; Gross, E.P., Phys. Rev. , 75, 1851-1864 (1949)
\bibitem{nesterenko} V.O. Nesterenko, W. Kleinig, and P.-G. Reinhard, Eur. Phys. J. D {\bf19}, 57 (2002).
\bibitem{xia} C. Xia, C. Yin, and V.V. Kresin, PRL {\bf102}, 156802 (2009)
\bibitem{vomfelde} A. vom Felde, J. Sprösser-Prou, J. Fink: Phys. Rev. B 40, 10181 (1989)
\bibitem{schnatterly} S.E.Schnatterly, Solid State Physics, {\bf34}, 275-358 (1979)
\bibitem{sonnichsen}  C. Sönnichsen, T. Franzl, T. Wilk, G. von Plessen and J Feldmann, New Journal of Physics, {\b4}, (2002)
\bibitem{ouyang} F. Ouyang, P.E. Batson, and M. Isaacson, Phys. Rev. B {\bf46}, 15421 (1992).
\bibitem{bystryi} R.G. Bystryi and I.V. Morozov, J. Phys. B: At. Mol. Opt. Phys. {\bf48}, 015401 (2015)
\bibitem{liang} H.-P. Liang, L.-J. Wan, C.-L. Bai, and L. Jiang, J. Phys. Chem. B {\bf109}, 7795-7800, (2005)
\bibitem{zhou} Zhou J. Phys. Chem. C 2008, 112, 20233–20240
\bibitem{kolwas} K. Kolwas and A. Derkachova, Opto-Electronics review {\bf18},4, 429–437 (2010)
\bibitem{si_Wpl} H. Dinigen, Z. Phys. 180, {\bf105} (1964).
\bibitem{si_Wpl2} H.R. Philipp, H. Ehrenreich, Semiconductors and Semimetals, Volume 3, (1967).
\bibitem{si_Wpl_Spl} J. E. Rowe, G. Margaritondo, and S. B. Christman, Phys. Rev. B {\bf15}, 2195 (1977).
\bibitem{zhouhaus} H.S. Zhou, I. Honma, H. Komiyama, and J. W. Haus, Phys. Rev. B {\bf50}, 12052 (1994)
\bibitem{zhouhaus2} H.S. Zhou, I. Honma, J. W. Haus, H. Saabe, H. Komiyama, Journal of Luminescence
{\bf70}, 21-34 (1996)
\bibitem{lee} J.-H. Lee, M.A. Mahmoud, V. Sitterle, J. Sitterle, and J. Carson Meredith, J. Am. Chem. Soc. {\bf131}, 14 (2009)
\bibitem{liebsch} A. Liebsch, Phys. Rev. B {\bf36} 7378 (1987)
\bibitem{ashkroft} N.W. Ashcroft, N.D. Mermin, Solid State Physics, New York: Holt, Rinehart and Winston (1976)

\end{thebibliography}
\bibliographystyle{gost705}

\end{document}